\documentclass[preprint,prd,nofootinbib,tightenlines,amsmath]{revtex4}
\usepackage{graphicx}
\usepackage{bm}

\begin{document}
\baselineskip=15pt \parskip=5pt

\vspace*{3em}

\title{Kaon Decay into Three Photons Revisited}

\author{Shu-Yu Ho}
\affiliation{Department of Physics and Center for Theoretical Sciences, \\
National Taiwan University, Taipei 106, Taiwan}

\author{Jusak Tandean}
\affiliation{Department of Physics and Center for Theoretical Sciences, \\
National Taiwan University, Taipei 106, Taiwan}

\date{\today $\vphantom{\bigg|_{\bigg|}^|}$}

\begin{abstract}
We evaluate the rare radiative kaon decays \,$K_{\rm L,S}\to3\gamma$.\,
Applying the requirements of gauge invariance and Bose symmetry, we derive a general form
of the decay amplitude, including both parity-conserving and parity-violating contributions.
We employ a chiral-Lagrangian approach combined with dimensional analysis arguments to estimate
the branching ratios of these decays in the standard model, obtaining values as large as
\,${\cal B}\bigl(K_{\rm L}^{}\to3\gamma\bigr)\sim 1\times10^{-14}$\, and
\,${\cal B}\bigl(K_{\rm S}^{}\to3\gamma\bigr)\sim 2\times10^{-17}$,\,
which exceed those found previously by a~few orders of magnitude.
Measurements on the branching ratios which are significantly larger than these numbers
would likely hint at the presence of new physics beyond the standard model.

\end{abstract}

\maketitle

\section{Introduction}

The rare kaon decays into three photons, \,$K_{\rm L}\to3\gamma$\, and
\,$K_{\rm S}\to3\gamma$,\, can occur in the absence of $CP$ violation.
One might then naively expect from the measured branching ratio
\,${\cal B}\bigl(K_{\rm L}\to2\gamma\bigr)\simeq$ $5.5\times10^{-4}$\,~\cite{pdg} that
\,${\cal B}\bigl(K_{\rm L}\to3\gamma\bigr)\sim\alpha_{\rm em}^{}\,
{\cal B}\bigl(K_{\rm L}\to2\gamma\bigr)\sim 4\times10^{-6}$.\,
However, this expectation is already too large in comparison to the result of the first
experimental search for the $3\gamma$ mode,
\,${\cal B}\bigl(K_{\rm L}\to3\gamma\bigr)<2.4\times10^{-7}$\,~\cite{pdg,Barr:1995rq}.
As for \,$K_{\rm S}\to3\gamma$,\, there is currently no experimental information available
on~it, but it is likely to be more suppressed than expected as well.

It turns out that the considerable smallness of the \,$K\to3\gamma$\, rates has to do
with the constraints imposed on the decay amplitude by gauge invariance and Bose
symmetry~\cite{Heiliger:1993ja}.
Gauge invariance implies that the total angular momentum $J$ of any two of the three photons
in the $3\gamma$ final-state cannot be zero, whereas Bose statistics forbids the photon
pair to have \,$J=1$.\,
Since each of the photon pairs must have \,$J\ge2$,\, the decay amplitude suffers from
a large number of angular momentum suppression factors.

The rates of \,$K_{\rm L,S}\to3\gamma$\, were roughly estimated a while ago
in Ref.~\cite{Heiliger:1993ja}.
The calculation was based on a simple model in which \,$K\to3\gamma$\, is assumed to proceed
from \,$K\to\pi^0\pi^0\gamma$\, with the $\pi^0$ pair immediately converting into a photon pair.
The resulting branching ratios are tiny,
\,${\cal B}\bigl(K_{\rm L}\to3\gamma\bigr)\sim3\times10^{-19}$\, and
\,${\cal B}\bigl(K_{\rm S}\to3\gamma\bigr)\sim5\times10^{-22}$\,~\cite{Heiliger:1993ja}.

Here we take another look at these rare decays, partly motivated by a new search for
\,$K_{\rm L}\to3\gamma$\, currently being performed in the E391a experiment at KEK~\cite{e391a}.
Since the estimate obtained in Ref.\,\cite{Heiliger:1993ja} resulted from only one contributing
diagram, it is possible that other contributions exist which can enhance the decay rate.
In this study we start with a general form of the \,$K\to3\gamma$\, amplitude
subject to the restrictions from gauge invariance and Bose statistics.
We then use a chiral-Lagrangian framework along with dimensional-analysis arguments to explore
the size of various important contributions.
This finally leads us to \,$K_{\rm L,S}\to3\gamma$\, rates which exceed those
previously estimated by a few orders of magnitude.

\section{Decay amplitudes and rates}

The decay \,$K\to3\gamma$\, being a weak transition, its amplitude ${\cal M}(K\to3\gamma)$
generally consists of separate terms describing the parity conserving (PC) and parity violating
(PV) components of the transition.
Accordingly, one can write ${\cal M}(K\to3\gamma)$  as a sum of their respective contributions,
\begin{eqnarray}
{\cal M}(K\to3\gamma)  \,\,=\,\,
{\cal M}_{\rm PC}^{K} \,+\, {\cal M}_{\rm PV}^{K} ~,
\end{eqnarray}
\vspace{-5ex}
\begin{eqnarray} \label{pc+pv}
{\cal M}_{\rm PC}^{K} \,\,=\,\, \varepsilon_{1\alpha}^* \varepsilon_{2\beta}^*
\varepsilon_{3\mu}^*\, M_{\rm PC}^{\alpha\beta\mu} ~, \hspace{5ex}
{\cal M}_{\rm PV}^{K} \,\,=\,\, \varepsilon_{1\alpha}^* \varepsilon_{2\beta}^*
\varepsilon_{3\mu}^*\, M_{\rm PV}^{\alpha\beta\mu} ~,
\end{eqnarray}
where $\varepsilon_{1,2,3}^{}$ are the polarization vectors of the photons.

To construct a general form of the \,$K\to3\gamma$\, amplitude, one follows a well-known
prescription.
The presence of photons implies that ${\cal M}(K\to3\gamma)$ has to be gauge invariant.
With three photons in the final state, Bose statistics dictates that the amplitude be
symmetric under interchange of any two of the photons.
These requirements must be satisfied by ${\cal M}_{\rm PC}$ and ${\cal M}_{\rm PV}$
separately.

We deal with $M_{\rm PV}$ first, as it turns out to be simpler than $M_{\rm PC}$ and is also
more relevant to~\,$K_{\rm L}\to3\gamma$,\, which is our main decay of interest,
relegating some details to Appendix~\ref{derivation}.
We also assume that the photons are on-shell.
Thus we get
\begin{eqnarray} \label{mpv}
M_{\rm PV}^{\alpha\beta\mu} &\,=\,&
\bigl(g^{\alpha\beta} z-k_2^\alpha k_1^\beta\bigr) \bigl(k_1^\mu\, y-k_2^\mu\, x\bigr)\,F(x,y,z)
\nonumber \\ && \!\! +\;
\bigl(g^{\beta\mu} y-k_3^\beta k_2^\mu\bigr)\bigl(k_2^\alpha\, x-k_3^\alpha\, z\bigr)\,F(z,x,y)
\nonumber \\ && \!\! +\;
\bigl(g^{\alpha\mu} x-k_3^\alpha k_1^\mu\bigr)\bigl(k_3^\beta\, z-k_1^\beta\, y\bigr)\,F(y,z,x)
\nonumber \\ && \!\! +\; \bigl[
g^{\alpha\beta}\,\bigl(k_1^\mu\, y-k_2^\mu\, x\bigr) +
g^{\beta\mu}\,\bigl(k_2^\alpha\, x-k_3^\alpha\, z\bigr)
\nonumber \\ && ~~~ +\,
g^{\alpha\mu}\,\bigl(k_3^\beta\, z-k_1^\beta\, y\bigr)
\,+\, k_3^\alpha k_1^\beta k_2^\mu - k_2^\alpha k_3^\beta k_1^\mu \bigr] \, G(x,y,z) ~,
\end{eqnarray}
where $k_{1,2,3}^{}$ are the momenta of the photons with polarizations $\varepsilon_{1,2,3}^{}$,
respectively,
\begin{eqnarray} \label{xyz}
x \,\,=\,\, k_1^{}\cdot k_3^{} ~, \hspace{5ex} y \,\,=\,\, k_2^{}\cdot k_3^{} ~,
\hspace{5ex} z \,\,=\,\, k_1^{}\cdot k_2^{} ~,
\end{eqnarray}
and the functions $F$ and $G$ must be free of kinematic singularities and satisfy the relations
\begin{eqnarray} \label{fg}
F(u,v,w) &\,=\,& -F(v,u,w) ~, \nonumber \\
G(u,v,w) \,\,=\,\, -G(v,u,w) &\,=\,& -G(w,v,u) \,\,=\,\, -G(u,w,v) ~,
\end{eqnarray}
with $u,v,w$ each being any one of the invariants \,$k_i^{}\cdot k_j^{}$.\,
This amplitude agrees with the one derived in Ref.~\cite{Dicus:1975cz} for
\,$\pi^0\to3\gamma$.\,

For the parity-conserving contribution, the form with the desired symmetry properties can be
expressed as
\begin{eqnarray} \label{mpc}
M_{\rm PC}^{\alpha\beta\mu} &\,=\,&
\bigl[ \bigl(g^{\alpha\beta}z-k_2^\alpha k_1^\beta \bigr)\, \epsilon^{\mu\rho\sigma\tau}\,
{\cal F}(x,y,z)
+ \bigl( g^{\beta\mu}y-k_3^\beta k_2^\mu\bigr)\,\epsilon^{\alpha\rho\sigma\tau}\,{\cal F}(z,x,y)
\nonumber \\ && ~ +
\bigl( g^{\alpha\mu}x-k_3^\alpha k_1^\mu \bigr)\,\epsilon^{\beta\rho\sigma\tau}\,{\cal F}(y,z,x)
\bigr] k_{1\rho}^{}k_{2\sigma}^{}k_{3\tau}^{}
\nonumber \\ && \!\! +\;
\bigl[ \bigl(k_2^\mu k_1^\tau-k_1^\mu k_2^\tau\bigr)\,
\epsilon^{\alpha\beta\rho\sigma}\, {\cal H}(x,y,z) +
\bigl(k_3^\alpha k_2^\rho-k_2^\alpha k_3^\rho\bigr)\, \epsilon^{\beta\mu\sigma\tau}\,
{\cal H}(z,x,y)
\nonumber \\ && ~~~ +
\bigl(k_1^\beta k_3^\sigma-k_3^\beta k_1^\sigma\bigr)\, \epsilon^{\alpha\mu\rho\tau}\,
{\cal H}(y,z,x) \bigr] k_{1\rho}^{}k_{2\sigma}^{}k_{3\tau}^{}
\nonumber \\ && \!\! +\; \mbox{$\frac{1}{3}$}
\bigl( g^{\alpha\beta} \epsilon^{\mu\rho\sigma\tau} + g^{\rho\sigma} \epsilon^{\alpha\beta\mu\tau}
+ g^{\beta\rho} \epsilon^{\alpha\mu\sigma\tau} - g^{\alpha\sigma} \epsilon^{\beta\mu\rho\tau}
\nonumber \\ && ~~~~ +\,
g^{\beta\mu} \epsilon^{\alpha\rho\sigma\tau} + g^{\sigma\tau} \epsilon^{\alpha\beta\mu\rho}
+ g^{\mu\sigma} \epsilon^{\alpha\beta\rho\tau} - g^{\beta\tau} \epsilon^{\alpha\mu\rho\sigma}
\nonumber \\ && ~~~~ +\,
g^{\alpha\mu} \epsilon^{\beta\rho\sigma\tau} + g^{\rho\tau} \epsilon^{\alpha\beta\mu\sigma}
+ g^{\mu\rho} \epsilon^{\alpha\beta\sigma\tau} - g^{\alpha\tau} \epsilon^{\beta\mu\rho\sigma}
\bigr) k_{1\rho}^{}k_{2\sigma}^{}k_{3\tau}^{}\, {\cal G}(x,y,z) ~,
\end{eqnarray}
where the functions $\cal F$, $\cal G$, and $\cal H$  are also free of kinematic
singularities and satisfy
\begin{eqnarray} \label{fgh}
{\cal F}(u,v,w) \,\,=\,\, -{\cal F}(v,u,w) ~, \hspace{1ex} && \hspace{3ex}
{\cal H}(u,v,w) \,\,=\,\, -{\cal H}(v,u,w) ~, \nonumber \\
{\cal G}(u,v,w) \,\,=\,\, -{\cal G}(v,u,w) &\,=\,& -{\cal G}(w,v,u) \,\,=\,\, -{\cal G}(u,w,v) ~.
\end{eqnarray}
The formula for $M_{\rm PC}$ above may have been constructed for the first time in this paper.

After summing \,$|{\cal M}_{\rm PV}^{K}+{\cal M}_{\rm PC}^{K}|^2$\, over
the photon polarizations, we find that there is no interference between the PC and PV
contributions in the result, in accord with expectation.
It is given~by
\begin{eqnarray} \label{M2}
\sum_{\rm pol}|{\cal M}(K\to3\gamma)|^2  \,\,=\,\,
\sum_{\rm pol}\Bigl( \bigl|{\cal M}_{\rm PV}^{K}\bigr|^2 \,+\,
\bigl|{\cal M}_{\rm PC}^{K}\bigr|^2 \Bigr) ~, ~~~~
\end{eqnarray}
where
\begin{eqnarray} \label{M2pv}
\sum_{\rm pol}\bigl|{\cal M}_{\rm PV}^{K}\bigr|^2 \,= \begin{array}[t]{l}
4\bigl\{ |F_1^{}|^2 z^2 + |F_2^{}|^2 y^2 + |F_3^{}|^2 x^2 \,+\, 2\,|G|^2
\vspace{1ex} \\ ~~ +\,
{\rm Re}\bigl[ F_1^*F_2^{}\,y\,z+F_2^*F_3^{}\,x\,y+F_3^*F_1^{}\,x\,z
+ 2\bigl(F_1^*z+F_2^*y+F_3^*x\bigr)G \bigr] \bigr\} x\,y\,z ~, ~~~~
\end{array}
\end{eqnarray}
\begin{eqnarray} \label{M2pc}
\sum_{\rm pol}\bigl|{\cal M}_{\rm PC}^{K}\bigr|^2 \,= \begin{array}[t]{l}
4\bigl\{\bigl(|{\cal F}_1^{}|^2 + |{\cal H}_1^{}|^2\bigr) z^2 +
\bigl(|{\cal F}_2^{}|^2+|{\cal H}_2^{}|^2\bigr) y^2 +
\bigl(|{\cal F}_3^{}|^2+|{\cal H}_3^{}|^2\bigr) x^2 \,+\, 2\,|{\cal G}|{}^2
\vspace{1ex} \\ ~~ +\,
{\rm Re}\bigl[
\bigl({\cal F}_1^*+{\cal H}_1^*\bigr)\bigl({\cal F}_2^{}+{\cal H}_2^{}+2{\cal G}/y\bigr) y\,z +
\bigl({\cal F}_2^*+{\cal H}_2^*\bigr)\bigl({\cal F}_3^{}+{\cal H}_3^{}+2{\cal G}/x\bigr) x\,y
\vspace{1ex} \\ \hspace*{8ex} +~
\bigl({\cal F}_3^*+{\cal H}_3^*\bigr)\bigl({\cal F}_1^{}+{\cal H}_1^{}+2{\cal G}/z\bigr) x\,z
\bigr] \bigr\} x\,y\,z ~, ~~~~
\end{array}
\end{eqnarray}
\begin{eqnarray} \label{f123}
F_1^{} \,\,=\,\, F(x,y,z) ~, \hspace{5ex} F_2^{} \,\,=\,\,  F(z,x,y) ~, \hspace{5ex}
F_3^{} \,\,=\,\, F(y,z,x) ~,
\end{eqnarray}
similarly for ${\cal F}_{1,2,3}^{}$ and ${\cal H}_{1,2,3}^{}$, and
\begin{eqnarray} \label{gg}
G \,\,=\,\, G(x,y,z) ~, \hspace{5ex} {\cal G} \,\,=\,\, {\cal G}(x,y,z) ~.
\end{eqnarray}
The resulting decay rate can be written as
\begin{eqnarray} \label{Gamma}
\Gamma(K\to3\gamma) \,\,=\,\, \frac{1}{256\,\pi^3\,m_K^3}\,\frac{1}{3!}
\int ds_{12}^{}\,ds_{23}^{} \sum_{\rm pol}|{\cal M}(K\to3\gamma)|^2 ~,
\end{eqnarray}
where the 3! accounts for the three photons being identical particles and
\,$s_{mn}^{}=(k_m^{}+k_n^{})^2$.\,
Using the formulas above, we provide our numerical estimates in the next section.

Before moving on, we remark that the expressions in Eqs.~(\ref{mpv}), (\ref{mpc}),
(\ref{M2pv}), and~(\ref{M2pc}) apply more generally to the decay of other neutral
pseudoscalar particle into three photons.
Furthermore, they also work for a neutral scalar particle decaying into three photons,
but with the PC and PV contributions interchanged.

\section{Estimate of \,$\bm{K_{\rm L,S}\to3\gamma}$\, rates\label{rates}}

To explore the size of the leading contributions to ${\cal M}(K\to3\gamma)$, we adopt
a chiral-Lagrangian approach~\cite{dgh}.
Accordingly, they are expected to arise from the relevant terms in the chiral expansion and
give rise to terms in the functions $F$, $G$, $\cal F$, $\cal G$, and $\cal H$
with the lowest numbers of powers of the photon momenta~$k_i^{}$.
Since there are in principle many possible contributions to the amplitude, from tree and
loop diagrams, with mostly unknown parameters, we consider a~few representative
contributions and rely on dimensional-analysis arguments to determine their size.

\subsection{$\bm{K_{\rm L}\to3\gamma}$}

Neglecting $CP$ violation, we can concentrate on the PV part of the amplitude,
in Eq.\,(\ref{mpv}), as~\,$K_{\rm L}\to3\gamma$\, violates charge-conjugation invariance.
Thus, for $F$ and $G$ satisfying Eq.\,(\ref{fg}) we find the simplest form
\begin{eqnarray}
&& F(u,v,w) \,\,=\,\, c_F^{}(u-v) ~, \nonumber \\
G(u,v,w) &=& c_G^{}[(u-v)f(w)+(v-w)f(u)+(w-u)f(v)] ~,
\end{eqnarray}
where $c_{F,G}^{}$ are constants and $f$ is any well-behaved function, although it cannot be
a constant if $G$ is to be nonzero.
This implies that $F$ and $G$ contain at least two and four powers of $k_i^{}$, respectively,
as the momentum power in the chiral expansion in the meson sector always increases by even
numbers.
It follows that $M_{\rm PV}$ in Eq.\,(\ref{mpv}) contains at least seven powers of $k_i^{}$.

To assess the leading contributions to $M_{\rm PV}$, we first consider an example of a weak
chiral Lagrangian for strangeness changing, \,$|\Delta S|=1$,\, transitions within
the standard model~(SM) which is odd under parity, has seven derivatives,
and couples a kaon to three photons in a gauge-invariant way.
As is well known, the weak chiral Lagrangian for such transitions in the SM is dominated by
contributions which transform as $\bigl(8_{\rm L},1_{\rm R}\bigr)$~\cite{dgh} and has to be
invariant under the $CPS$ transformation~\cite{Bernard:1985wf}, which is the product
of the ordinary $CP$ transformation and the switching of the $s$ and $d$ quarks.
An example with the required properties is
\begin{eqnarray} \label{Lpv7}
{\cal L}_{\rm PV}^{} &=&
c_7^{}\, \bigl\langle \xi^\dagger h\xi\, \bigl(\nabla^\alpha{\cal V}^{\mu\nu}\bigr)
\bigl[ {\cal U}^\rho\, \nabla_{\!\alpha}^{}{\cal V}_{\rho\sigma}^{} +
\bigl(\nabla_\sigma^{}{\cal V}_{\rho\alpha}^{}\bigr) {\cal U}^\rho \bigr]\,
\nabla^\sigma{\cal V}_{\mu\nu}^{} \bigr\rangle
\;+\; {\rm H.c.}
\nonumber \\ &=&
\frac{8\sqrt2\,c_7^{}\,e^3}{27\,f_\pi^{}}\, \partial^\alpha F^{\mu\nu}\,
\bigl(\partial_{\!\alpha}^{}F_{\rho\sigma}^{}+\partial_\sigma^{}F_{\rho\alpha}^{}\bigr)\,
\partial^\rho\bar K^0\, \partial^\sigma F_{\mu\nu}^{} \;+\; \cdots \;+\; {\rm H.c.} ~,
\end{eqnarray}
where $c_7^{}$ is a constant, \,$F_{\mu\nu}=\partial_\mu A_\nu-\partial_\nu A_\mu$\, is the usual
photon field strength tensor, and only the relevant part is displayed in the second line.
In the first line, \,$\xi=e^{i\varphi/(2f)}$\, contains the lightest octet of pseudoscalar mesons
via the 3$\times$3 matrix~$\varphi$~\cite{dgh}, with \,$f=f_\pi^{}=92$\,MeV\, being the pion
decay constant, $h$ is a 3$\times$3 matrix having elements
\,$h_{kl}^{}=\delta_{k2}^{}\delta_{3l}^{}$\, which selects out \,$s\to d$\, transitions,
\,${\cal V}_{\mu\nu}=e\bigl(\xi^\dagger Q\xi+\xi Q\xi^\dagger\bigr)F_{\mu\nu}$\, with
\,$Q={\rm diag}(2,-1,-1)/3$\, being the quark charge matrix,
\,$\nabla_\alpha X=\partial_\alpha X+\frac{1}{2}\bigl[
\xi^\dagger\,\partial_\alpha\xi+\xi\, \partial_\alpha\xi^\dagger
+ ie\bigl(\xi Q\xi^\dagger+\xi^\dagger Q \xi\bigr)A_\alpha,X\bigr]$,\,
and
\,${\cal U}_\alpha=i\bigl(\xi^\dagger\,\partial_\alpha\xi-\xi\,\partial_\alpha\xi^\dagger\bigr)
+e\bigl(\xi Q\xi^\dagger-\xi^\dagger Q\xi\bigr)A_\alpha$.\,
The Lagrangian in Eq.\,(\ref{Lpv7}) yields
\begin{eqnarray} \label{fg7}
F(u,v,w) \,\,=\,\, \frac{32\sqrt2\;i c_7^{}\,e^3}{27\,f_\pi^{}}(u-v) ~, \hspace{5ex}
G(u,v,w) \,\,=\,\, 0
\end{eqnarray}
in ${\cal M}\bigl(\bar K^0\bigr)$ and the same functions in ${\cal M}\bigl(K^0\bigr)$,
but with $c_7^{}$ replaced with~$c_7^*$ if $CP$ violation is present.

If $CP$ is conserved, only ${\cal M}_{\rm PV}$ contributes to the \,$K_{\rm L}^{}\to3\gamma$\,
amplitude.
In that case, upon applying Eq.\,(\ref{fg7}) in Eq.\,(\ref{M2pv}) and adopting the convention
\,$K_{\rm L}^{}=\bigl(K^0+\bar K^0\bigr)/\sqrt2$,\, we find
\begin{eqnarray} \label{M7}
\sum_{\rm pol}\bigl|{\cal M}\bigl(K_{\rm L}^{}\to3\gamma\bigr)\bigr|^2 \,\,=\,\,
\frac{|128\,c_7^{}|^2\,e^6}{729\,f_\pi^2}
\bigl(x^2 y^2+y^2 z^2+x^2 z^2-x\,y\,z^2-x\,y^2\,z-x^2\,y\,z\bigr)x\,y\,z ~,
\end{eqnarray}
having made use of Eq.\,(\ref{xyz}).
To calculate the decay rate, we then need the value of $c_7^{}$.
Since it is not possible at present to determine this constant rigorously from
the quark-level parameters, we estimate it with the aid of naive dimensional
analysis~\cite{Manohar:1983md}.
Thus we obtain the order-of-magnitude value
\begin{eqnarray} \label{c7}
c_7^{} \,\,\sim\,\,
\frac{G_{\rm F}^{}\, \lambda_{\rm C}^{}\,f_\pi^4}{\sqrt2\,\Lambda^8}
\,\,\simeq\,\, 1.0\times 10^{-9} {\rm~GeV}^{-6} ~,
\end{eqnarray}
where \,$\lambda_{\rm C}^{}=0.22$\, is the Cabibbo mixing parameter and
\,$\Lambda$ represents the scale at which the chiral Lagrangian approach breaks down,
and so we take \,$\Lambda=m_\rho^{}=775$\,MeV\,~\cite{pdg}.
The resulting branching ratio is
\begin{eqnarray} \label{b7(kl)}
{\cal B}\bigl(K_{\rm L}^{}\to3\gamma\bigr) \,\,\sim\,\, 7.4\times10^{-17} ~.
\end{eqnarray}

This number is about 250 times larger than the earlier prediction,
\,${\cal B}\bigl(K_{\rm L}\to3\gamma\bigr)\sim3\times10^{-19}$,\, given in
Ref.~\cite{Heiliger:1993ja}.
We have repeated that calculation and come to a different result,
which we briefly discuss here.
In the model employed in Ref.~\cite{Heiliger:1993ja}, the amplitude
${\cal M}\bigl(K_{\rm L}^{}\to3\gamma\bigr)$ is roughly approximated by
\begin{eqnarray} \label{imM}
{\rm Im}\,{\cal M}\bigl(K_{\rm L}^{}\to3\gamma\bigr) &=& \theta\bigl(s_{12}^{}-4m_\pi^2\bigr)\,
\mbox{$\frac{1}{2}$}\int\frac{d^3p_1^{}}{(2\pi)^3\,2p_{10}^{}}
\frac{d^3p_2^{}}{(2\pi)^3\,2 p_{20}^{}}\;
(2\pi)^4\,\delta^{(4)}\bigl(p_K^{}-p_1^{}-p_2^{}-k_3^{}\bigr)
\nonumber \\ && \times\;
\frac{h_{\rm L}^{}}{m_K^5} \bigl(\varepsilon_3^*\!\cdot\!p_1^{}\,k_3^{}\!\cdot\!p_K^{}
- \varepsilon_3^*\!\cdot\!p_K^{}\,k_3^{}\!\cdot\!p_1^{}\bigr)\,
k_3^{}\!\cdot\!\bigl(2p_1^{}-p_K^{}\bigr)
\nonumber \\ && \times\;
\frac{\tilde G\,s_{12}^{}}{m_\rho^2} \Biggl(
\frac{k_1^{}\!\cdot\!p_1^{}\,k_2^{}\!\cdot\!p_1^{}}{k_1^{}\!\cdot\!k_2^{}}\,g_{\mu\nu}^{}
+ p_{1\mu}^{}p_{1\nu}^{}
- \frac{k_1^{}\!\cdot\!p_1^{}}{k_1^{}\!\cdot\!k_2^{}}\,k_{2\mu}^{}p_{1\nu}^{} -
\frac{k_2^{}\!\cdot\!p_1^{}}{k_1^{}\!\cdot\!k_2^{}}\,k_{1\nu}^{}p_{1\mu}^{} \Biggr)
\varepsilon_1^{*\mu} \varepsilon_2^{*\nu}
\nonumber \\ && \!\!\! +\;
\bigl[\mbox{permutations of $\bigl(\varepsilon_1^{},k_1^{}\bigr)$,
$\bigl(\varepsilon_2^{},k_2^{}\bigr)$, $\bigl(\varepsilon_3^{},k_3^{}\bigr)$}\bigr] ~,
\end{eqnarray}
coming from a $\pi^0$-loop diagram, where $p_K^{}$ is the $K_{\rm L}$ momentum,
$p_{1,2}^{}$ are the $\pi^0$ momenta in the loop, \,$h_{\rm L}^{}\simeq6.6\times10^{-8}$,\,
and \,$\tilde G=\frac{10}{9}\,g_{\omega\pi\gamma}^2$\,
with \,$g_{\omega\pi\gamma}^{}=0.77{\rm\,GeV}^{-1}$.\,
To compute this requires the evaluation of the integrals
\begin{eqnarray} \label{ints}
K^{\mu\nu\rho\sigma} &=& \int\frac{d^3p_1^{}}{2p_{10}^{}} \frac{d^3p_2^{}}{2p_{20}^{}}\;
\delta^{(4)}\bigl(P-p_1^{}-p_2^{}\bigr)\; p_1^\mu p_1^\nu p_1^\rho p_1^\sigma\,
f\bigl(p_1^{}\cdot p_2^{}\bigr) ~,
\nonumber \\
L^{\mu\nu\rho\sigma} &=& \int\frac{d^3p_1^{}}{2p_{10}^{}} \frac{d^3p_2^{}}{2p_{20}^{}}\;
\delta^{(4)}\bigl(P-p_1^{}-p_2^{}\bigr)\; p_1^\mu p_1^\nu p_1^\rho p_2^\sigma\,
f\bigl(p_1^{}\cdot p_2^{}\bigr) ~.
\end{eqnarray}
We have collected the results in Appendix~\ref{integrals}, where the expression for
$K^{\mu\nu\rho\sigma}$ agrees with that given in Ref.~\cite{Heiliger:1993ja},
but our $L^{\mu\nu\rho\sigma}$ differs from theirs.
With \,${\rm Im}\,{\cal M}\bigl(K_{\rm L}^{}\to3\gamma\bigr)$\, computed and subsequently
equated to
\,${\cal M}_{\rm PV}\bigl(K_{\rm L}^{}\bigr)=
\varepsilon_{1\alpha}^*\varepsilon_{2\beta}^*\varepsilon_{3\mu}^*M_{\rm PV}^{\alpha\beta\mu}$,\,
we then extract
\begin{eqnarray} \label{fghms}
F(u,v,w) &=& c\,
\biggl[ \frac{1}{\sqrt v}\bigl(v-2m_\pi^2\bigr)^{\!5/2}\,\theta\bigl(v-2m_\pi^2\bigr)
\,-\, \frac{1}{\sqrt u}\bigl(u-2m_\pi^2\bigr)^{\!5/2}\,\theta\bigl(u-2m_\pi^2\bigr) \biggr] ~,
\nonumber \\
G(u,v,w) &=&  \frac{c}{2} \biggl[
\frac{w-v}{\sqrt u}\bigl(u-2m_\pi^2\bigr)^{\!5/2}\,\,\theta\bigl(u-2m_\pi^2\bigr) \,+\,
\frac{u-w}{\sqrt v}\bigl(v-2m_\pi^2\bigr)^{\!5/2}\,\theta\bigl(v-2m_\pi^2\bigr)
\nonumber \\ && \hspace*{3ex} +\;
\frac{v-u}{\sqrt w}\bigl(w-2m_\pi^2\bigr)^{\!5/2}\,\theta\bigl(w-2m_\pi^2\bigr)
\biggr] ~, ~~~~~~
\\ \nonumber \\
\label{chms}
c &=& \frac{\tilde G\,h_{\rm L}^{}}{120\,\pi\,m_K^5\,m_\rho^2}
\,\,\simeq\,\,  6.3\times 10^{-9} {\rm~GeV}^{-9} ~.
\end{eqnarray}
Incorporating these into Eqs.~(\ref{M2pv}) and~(\ref{Gamma}) leads us to
\,${\cal B}\bigl(K_{\rm L}\to3\gamma\bigr)\sim1.0\times10^{-18}$,\,
which is three times greater than the number found in~Ref.~\cite{Heiliger:1993ja}.

Since the results in the preceding paragraph arise from a loop diagram~\cite{Heiliger:1993ja},
it is instructive to compare them to tree-level contributions of the same chiral order.
From the $F$ and $G$ formulas in Eq.\,(\ref{fghms}) we can see that they contain four and
six powers of the photon momenta~$k_i^{}$, respectively, and hence their contributions
to ${\cal M}\bigl(K_{\rm L}^{}\to3\gamma\bigr)$ have nine powers of $k_i^{}$.
If a weak chiral Lagrangian with nine derivatives contributes to this amplitude at tree level,
the size of its coupling constant in $F(u,v,w)$ is predicted by naive dimensional analysis to be
\begin{eqnarray}
c' \,\,\sim\,\,
\frac{e^3\,G_{\rm F}^{}\, \lambda_{\rm C}^{}\,f_\pi^3}{\sqrt2\,\Lambda_{\vphantom{|}}^{\!10}}
\,\,\sim\,\, 5.0\times 10^{-10} {\rm~GeV}^{-9} ~.
\end{eqnarray}
This is about 12~times smaller than $c$ in Eq.\,(\ref{chms}).
One may then suggest based on this comparison that loop contributions with seven powers
of $k_i^{}$ could also be enhanced by a similar amount relative to the tree-level
contribution from Lagrangians with seven derivatives, such as that in Eq.\,(\ref{Lpv7}).
If~such enhancement occurs, we may have
\begin{eqnarray} \label{b7(kl)'}
{\cal B}\bigl(K_{\rm L}^{}\to3\gamma\bigr) \,\,\sim\,\, 1\times10^{-14} ~,
\end{eqnarray}
instead of Eq.\,(\ref{b7(kl)}).
The numbers in Eqs.~(\ref{b7(kl)}) and~(\ref{b7(kl)'}) can be taken to be representative
values coming from the various contributions to this decay
and also to indicate the level of uncertainty involved in our crude calculation.
It is therefore reasonable to conclude that
\begin{eqnarray} \label{b7(kl)range}
7\times10^{-17} \,\,\lesssim\,\, {\cal B}\bigl(K_{\rm L}^{}\to3\gamma\bigr)
\,\,\lesssim\,\, 1\times10^{-14} ~.
\end{eqnarray}

If $CP$ violation is not neglected, then ${\cal M}_{\rm PC}^{K}$ also contributes
to the  \,$K_{\rm L}^{}\to3\gamma$\,  rate, via the $\epsilon_K^{}$ term in
\,$K_{\rm L}^{}\simeq\bigl[K^0+\bar K^0+\epsilon_K^{}\bigl(K^0-\bar K^0\bigr)\bigr]/\sqrt2$,\,
where \,$\epsilon_K^{}\sim2\times10^{-3}$\, is the $CP$-violation parameter in kaon mixing.
Now, the similarity between Eqs.~(\ref{fg}) and~(\ref{fgh}) suggests that
$\cal F$ and $\cal H$ ($\cal G$) are comparable in size to $F$ ($G$).
We can then expect that ${\cal M}_{\rm PC}^{K}$ is, at most, also comparable to
${\cal M}_{\rm PV}^{K}$.
This implies that the effect of $CP$ violation on
\,${\cal B}\bigl(K_{\rm L}^{}\to3\gamma\bigr)$\, is small,
and hence the predicted branching ratio is still what is quoted in~Eq.\,(\ref{b7(kl)range}).

\subsection{$\bm{K_{\rm S}\to3\gamma}$}

The rate of this decay is determined mostly by the parity-conserving contribution
${\cal M}_{\rm PC}^{K}$.
The branching ratio predicted in Ref.~\cite{Heiliger:1993ja} is
\,${\cal B}\bigl(K_{\rm S}\to3\gamma\bigr)\sim5\times10^{-22}$.\,
Repeating the calculation, as in the \,$K_{\rm L}\to3\gamma$\, case, we find instead
a value three times larger,
\,${\cal B}\bigl(K_{\rm S}\to3\gamma\bigr)\sim 1.8\times10^{-21}$.\,
Since this arises from a loop contribution involving nine powers of the photon momenta $k_i^{}$,
we need to consider as before the lower-order contributions, with seven powers of $k_i^{}$,
which may be larger.

We take the leading-order form
\,${\cal F}(u,v,w)\sim{\cal H}(u,v,w)=\tilde c(u-v)$,\,
satisfying Eq.\,(\ref{fgh}), with $\tilde c$ being a constant and \,${\cal G}=0$.\,
In this case, the situation is similar to that of ${\cal M}_{\rm PC}^{K}$ with $F$ and
$G$ given in Eq.\,(\ref{fg7}).
More precisely, making a comparison of
\,{\footnotesize$\sum$}$_{\rm pol}^{}|{\cal M}_{\rm PC}^{K}|^2$\, in
Eq.\,(\ref{M2pc}) and \,{\footnotesize$\sum$}$_{\rm pol}^{}|{\cal M}_{\rm PV}^{K}|^2$\, in
Eq.\,(\ref{M2pv}) for the two cases, respectively, one can see that their decay distributions
have the same functional dependence on~$x$, $y$, and $z$.
It follows that $\Gamma\bigl(K_{\rm S}^{}\to3\gamma\bigr)$ can be expected to be roughly of
the same order as $\Gamma\bigl(K_{\rm L}^{}\to3\gamma\bigr)$.
Interestingly, the measured rates of the corresponding $2\gamma$ modes are also of similar order,
\,$\Gamma\bigl(K_{\rm S}^{}\to2\gamma\bigr)\sim
2.7\,\Gamma\bigl(K_{\rm L}^{}\to2\gamma\bigr)$\,~\cite{pdg}.
In~view of Eq.\,(\ref{b7(kl)range}), we can therefore predict that
\begin{eqnarray} \label{b7(ks)}
1\times10^{-19} \,\,\lesssim\,\, {\cal B}\bigl(K_{\rm S}^{}\to3\gamma\bigr)
\,\,\lesssim\,\, 2\times10^{-17} ~.
\end{eqnarray}

\section{Conclusions\label{concl}}

We have revisited the rare kaon decay \,$K\to3\gamma$,\, which is expected to be much
suppressed because its amplitude has a large number of angular momentum suppression factors.
We have constructed a general form of the decay amplitude which satisfies
the requirements of gauge invariance and Bose symmetry and
includes both parity-conserving and parity-violating contributions.
We have in addition calculated the squared amplitude, summed over the photon polarizations,
which can be useful to produce a Dalitz plot distribution of the decay.
These results are applicable generally to the decay of any spinless particle into
three photons.

More specifically, we have dealt mainly with \,$K_{\rm L}\to3\gamma$,\, which is currently
the subject of a~new experimental search at KEK, but also evaluated~\,$K_{\rm S}\to3\gamma$,\,
albeit more briefly.
To explore the leading-order contributions to their amplitudes,
we have adopted a~chiral-Lagrangian approach in the context of the standard model.
This implies that there are many possible contributions to the amplitudes, from tree
and loop diagrams, with mostly unknown parameters.
Consequently, for definiteness we have considered a~number of representative
contributions and used dimensional-analysis arguments to estimate their size.
This has finally led us to arrive at branching ratios as large as
\,${\cal B}\bigl(K_{\rm L}^{}\to3\gamma\bigr)\sim 1\times10^{-14}$\, and
\,${\cal B}\bigl(K_{\rm S}^{}\to3\gamma\bigr)\sim 2\times10^{-17}$,\,
which exceed those estimated before by a~few orders of magnitude, but are still very small.
Nevertheless, any experimental findings on the branching ratios which are significantly
greater than these numbers would likely signal the effect of new physics beyond
the standard model.

\acknowledgments

This work was supported in part by NSC and NCTS.
We would like to thank Y.~Bob~Hsiung and Yu-Chen Tung for discussions and experimental
information.
We also thank X.G.~He and G.~Valencia for valuable comments.

\appendix

\section{Derivation of amplitudes\label{derivation}}

To obtain a general form of the \,$K\to3\gamma$\, amplitude with the desired properties,
we start with expressions for $M_{\rm PV}^{\alpha\beta\mu}$ and $M_{\rm PC}^{\alpha\beta\mu}$
consisting of terms involving all possible combinations of the available tensors,
$k_{1,2,3}^\alpha$, $g^{\eta\kappa}$, and $\epsilon^{\mu\nu\sigma\tau}$.
Thus we have
\begin{eqnarray}
M_{\rm PV}^{\alpha\beta\mu} &=&
g^{\alpha\beta}\, \bigl( k_1^\mu\, a_1^{} + k_2^\mu\, a_2^{} \bigr) +
g^{\alpha\mu}\, \bigl( k_1^\beta\, a_3^{} + k_3^\beta\, a_4^{} \bigr) +
g^{\beta\mu}\, \bigl( k_2^\alpha\, a_5^{} + k_3^\alpha\, a_6^{} \bigr)
\\ && \!\! +\;  \nonumber
k_2^\alpha \bigl[ k_1^\beta \bigl(k_1^\mu b_1^{} + k_2^\mu b_2^{}\bigr) +
k_3^\beta \bigl(k_1^\mu b_3^{} + k_2^\mu b_4^{}\bigr) \bigr] +
k_3^\alpha \bigl[ k_1^\beta \bigl(k_1^\mu b_5^{} + k_2^\mu b_6^{}\bigr) +
k_3^\beta \bigl(k_1^\mu b_7^{} + k_2^\mu b_8^{}\bigr) \bigr] ~, ~~~~
\end{eqnarray}
\begin{eqnarray}
M_{\rm PC}^{\alpha\beta\mu} &=&
\epsilon^{\alpha\beta\mu\rho}\,
\bigl( k_{1\rho}^{} c_1^{} + k_{2\rho}^{} c_2^{} + k_{3\rho}^{} c_3^{} \bigr) +
\bigl( g^{\alpha\beta} \epsilon^{\mu\rho\sigma\tau} d_1^{}
+ g^{\alpha\mu} \epsilon^{\beta\rho\sigma\tau} d_2^{}
+ g^{\beta\mu}\epsilon^{\alpha\rho\sigma\tau}d_3^{}\bigr) k_{1\rho}^{}k_{2\sigma}^{}k_{3\tau}^{}
\nonumber \\ && \!\! +\; \bigl[
\bigl(k_2^\alpha\, f_1^{}+ k_3^\alpha\, f_2^{}\bigr) \epsilon^{\beta\mu\rho\sigma} +
\bigl(k_1^\beta\, f_3^{} + k_3^\beta\, f_4^{}\bigr) \epsilon^{\alpha\mu\rho\sigma} +
\bigl(k_1^\mu\, f_5^{} + k_2^\mu\, f_6^{}\bigr) \epsilon^{\alpha\beta\rho\sigma}
\bigr] k_{1\rho}^{}k_{2\sigma}^{}
\nonumber \\ && \!\! +\; \bigl[
\bigl(k_3^\alpha\, g_1^{} + k_2^\alpha\, g_2^{}\bigr) \epsilon^{\beta\mu\rho\sigma} +
\bigl(k_1^\mu\, g_3^{} + k_2^\mu\, g_4^{}\bigr) \epsilon^{\alpha\beta\rho\sigma} +
\bigl(k_1^\beta\, g_5^{} + k_3^\beta\, g_6^{}\bigr) \epsilon^{\alpha\mu\rho\sigma}
\bigr] k_{1\rho}^{}k_{3\sigma}^{}
\nonumber \\ && \!\! +\; \bigl[
\bigl(k_2^\mu\, h_1^{} + k_1^\mu\, h_2^{}\bigr) \epsilon^{\alpha\beta\rho\sigma} +
\bigl(k_3^\beta\, h_3^{} + k_1^\beta\, h_4^{}\bigr) \epsilon^{\alpha\mu\rho\sigma} +
\bigl(k_3^\alpha\, h_5^{} + k_2^\alpha\, h_6^{}\bigr) \epsilon^{\beta\mu\rho\sigma}
\bigr] k_{2\rho}^{}k_{3\sigma}^{}
\nonumber \\ && \!\! +\;
\bigl\{ \bigl[ k_2^\alpha \bigl(k_1^\beta l_1^{}+k_3^\beta l_2^{}\bigr) +
k_3^\alpha \bigl(k_1^\beta l_3^{}+k_3^\beta l_4^{}\bigr) \bigr] \epsilon^{\mu\rho\sigma\tau} +
\bigl[ k_2^\alpha \bigl(k_1^\mu l_5^{}+k_2^\mu l_6^{}\bigr) +
k_3^\alpha \bigl(k_1^\mu l_7^{}+k_2^\mu l_8^{}\bigr) \bigr] \epsilon^{\beta\rho\sigma\tau}
\nonumber \\ && ~~~ +
\bigl[ k_1^\beta \bigl(k_1^\mu l_9^{}+k_2^\mu l_{10}^{}\bigr) +
k_3^\beta \bigl(k_1^\mu l_{11}^{}+k_2^\mu l_{12}^{}\bigr) \bigr]
\epsilon^{\alpha\rho\sigma\tau} \bigr\} k_{1\rho}^{}k_{2\sigma}^{}k_{3\tau}^{} ~,
\end{eqnarray}
where the $a_i^{}$, $b_i^{}$, $c_i^{}$, $d_i^{}$, $f_i^{}$, $g_i^{}$, $h_i^{}$,
and $l_i^{}$ are functions dependent on the invariants \,$k_i^{}\cdot k_j^{}$.\,
After the requirements of gauge invariance and Bose symmetry have been imposed on
$M_{\rm PV}^{\alpha\beta\mu}$ and $M_{\rm PC}^{\alpha\beta\mu}$  separately,
they each contain a much smaller number of functions.
For $M_{\rm PC}^{\alpha\beta\mu}$, we make further simplification with the aid of Schouten's
identity, which states that in four dimensions a tensor with five or more Lorentz indices
vanishes identically if it is completely antisymmetric with respect to five or more of
the indices.\footnote{Some other examples of the use of Schouten's identity can be found
in Ref.~\cite{Fearing:1994ga}.}
Such a tensor is
\begin{eqnarray} \label{sch}
g^{\alpha\mu} \epsilon^{\nu\rho\sigma\tau} - g^{\alpha\nu} \epsilon^{\mu\rho\sigma\tau}
- g^{\alpha\rho} \epsilon^{\nu\mu\sigma\tau} - g^{\alpha\sigma} \epsilon^{\nu\rho\mu\tau}
- g^{\alpha\tau} \epsilon^{\nu\rho\sigma\mu} \,\,=\,\, 0 ~,
\end{eqnarray}
which is fully antisymmetric with respect to \,$\mu,\nu,\rho,\sigma,\tau$.\,
We display our results for $M_{\rm PV}$ and $M_{\rm PC}$
in Eqs.~(\ref{mpv}) and~(\ref{mpc}), respectively, in the case of on-shell photons.
The $\cal G$ terms in Eq.\,(\ref{mpc}) could be simplified further using~Eq.\,(\ref{sch}),
but then they would not be manifestly Bose-symmetric.

\section{Integrals\label{integrals}}

The integrals in Eq.\,(\ref{ints}) can be written in terms of all the possible appropriate
combinations of the available tensors, $g^{\alpha\beta}$ and $P^\eta$,~as
\begin{eqnarray}
K^{\mu\nu\rho\sigma} &=&
\frac{\pi\,\lambda^{1\over2}\bigl(s,m_1^2,m_2^2\bigr)}{10\,s} \Bigl[
\bigl(g^{\mu\nu}g^{\rho\sigma}+g^{\mu\rho}g^{\nu\sigma}+g^{\mu\sigma}g^{\nu\rho}\bigr)K_1^{}
+ P^\mu P^\nu P^\rho P^\sigma\,K_3^{}
\nonumber \\ && \hspace*{17ex} +\;
\bigl(g^{\mu\nu}P^\rho P^\sigma+g^{\mu\rho}P^\nu P^\sigma+g^{\mu\sigma}P^\nu P^\rho
+ g^{\nu\rho}P^\mu P^\sigma
\\ && \hspace*{21ex} +\;
g^{\nu\sigma}P^\mu P^\rho + g^{\rho\sigma}P^\mu P^\nu\bigr) K_2^{} \Bigr]
f\Bigl(\mbox{$\frac{1}{2}$}\bigl(s-m_1^2-m_2^2\bigr)\Bigr) ~, \nonumber \
\end{eqnarray}
\begin{eqnarray}
L^{\mu\nu\rho\sigma} &=&
\frac{\pi\,\lambda^{1\over2}\bigl(s,m_1^2,m_2^2\bigr)}{10\,s} \Bigl[
\bigl(g^{\mu\nu}g^{\rho\sigma}+g^{\mu\rho}g^{\nu\sigma}+g^{\nu\rho}g^{\mu\sigma}\bigr) L_1^{}
+ P^\mu P^\nu P^\rho P^\sigma\, L_4^{}
\nonumber \\ && \hspace*{17ex} +\;
\bigl(g^{\mu\nu}P^\rho P^\sigma+g^{\mu\rho}P^\nu P^\sigma+g^{\nu\rho}P^\nu P^\sigma\bigr) L_2^{}
\\ && \hspace*{17ex} +\;
\bigl(g^{\mu\sigma}P^\nu P^\rho+g^{\nu\sigma}P^\mu P^\rho+g^{\rho\sigma}P^\mu P^\nu\bigr) L_3^{}
\Bigr] f\Bigl(\mbox{$\frac{1}{2}$}\bigl(s-m_1^2-m_2^2\bigr)\Bigr) ~, ~~~~ \nonumber
\end{eqnarray}
where \,$\lambda(u,v,w)=u^2+v^2+w^2-2u\,v-2v\,w-2u\,w$,\, \,$m_1^2=p_1^2$,\, \,$m_2^2=p_2^2$,\,
and the coefficients $K_{1,2,3}^{}$ and $L_{1,2,3,4}^{}$ are functions of \,$s=P^2$\, and
\,$m_{1,2}^{}$.\,
We then derive
\begin{eqnarray}
K_1^{} &=& \frac{1}{48\,s^2}
\bigl[s^2-2s\,\bigl(m_1^2+m_2^2\bigr)+\bigl(m_1^2-m_2^2\bigr){}^2\bigr]^2 \,,
\nonumber \\ \nonumber \\
K_2^{} &=& \frac{-1}{24\,s^3}
\bigl[s-\bigl(m_1^{}+m_2^{}\bigr){}^2\bigr] \bigl[s-\bigl(m_1^{}-m_2^{}\bigr){}^2\bigr]
\bigl[3 s^2+s\,\bigl(4 m_1^2-6 m_2^2\bigr)+3\bigl(m_1^2-m_2^2\bigr){}^2\bigr] \,,
\nonumber \\ \nonumber \\
K_3^{} &=& \frac{1}{s^4}
\bigl\{s^4+ s^2\bigl(m_1^4-6 m_1^2 m_2^2+6 m_2^4\bigr)
+ s\bigl(m_1^2-4m_2^2\bigr)\bigl[s^2+\bigl(m_1^2-m_2^2\bigr){}^2\bigr]
+ \bigl(m_1^2-m_2^2\bigr){}^4\bigr\} \,, \nonumber \\
\end{eqnarray}
\begin{eqnarray}
L_1^{} &=& \frac{-1}{48\,s^2}
\bigl[s^2-2s\,\bigl(m_1^2+m_2^2\bigr)+\bigl(m_1^2-m_2^2\bigr){}^2\bigr]^2 \,,
\nonumber \\ \nonumber \\
L_2^{} &=& \frac{-1}{24\,s^3}
\bigl[s-\bigl(m_1^{}+m_2^{}\bigr){}^2\bigr] \bigl[s-\bigl(m_1^{}-m_2^{}\bigr){}^2\bigr]
\bigl[2 s^2+s\,\bigl(m_1^2+m_2^2\bigr)-3\bigl(m_1^2-m_2^2\bigr){}^2\bigr] \,, \hspace*{5ex}
\nonumber \\ \nonumber
\end{eqnarray}
\begin{eqnarray}
L_3^{} &=& \frac{1}{24\,s^3}
\bigl[s-\bigl(m_1^{}+m_2^{}\bigr){}^2\bigr] \bigl[s-\bigl(m_1^{}-m_2^{}\bigr){}^2\bigr]
\bigl[3 s^2+s\,\bigl(4 m_1^2-6 m_2^2\bigr)+3\bigl(m_1^2-m_2^2\bigr){}^2\bigr] \,,
\nonumber \\ \nonumber \\
L_4^{} &=& \frac{1}{4\,s^4}
\bigl[s^4+s^3\bigl(m_1^2+m_2^2\bigr) + s^2\bigl(m_1^4+4 m_1^2 m_2^2-9 m_2^4\bigr)
\nonumber \\ && \hspace*{5ex} +\,
s\bigl(m_1^2+11 m_2^2\bigr)\bigl(m_1^2-m_2^2\bigr){}^2
- 4 \bigl(m_1^2-m_2^2\bigr){}^4\bigr] \,.
\end{eqnarray}
Our formula above for $K^{\mu\nu\rho\sigma}$ $\bigl(L^{\mu\nu\rho\sigma}\bigr)$ agrees
(disagrees) with that found in Ref.~\cite{Heiliger:1993ja}.
In obtaining Eq.~(\ref{fghms}), we set \,$m_1^{}=m_2^{}=m_\pi^{}$.\,

\end{document}